\documentstyle[12pt]{article} \voffset=-30mm
\begin{document}
%-----------------------------------------------------------------------------
\def\be{\begin{equation} }
\def\ee{\end{equation} }
\def\p{\prime}
\def\a{\alpha}
\def\d{\delta}
\def\P{\prod\limits_{m \in I_{k}^{r}}}
\def\PQ{\prod\limits_{m=1}^{n}}
\def\v{\varepsilon}
%-----------------------------------------------------------------------------
\begin{titlepage}
%-----------------------------------------------------------------------------
\vspace{35mm}
\begin{center}
{\Large\bf {On the Kinetics of Multi-dimensional Fragmentation}}
\end{center}
\vspace{5mm}
\begin{center}
{P. Singh$^{\rm a}$ and M. K. Hassan$^{\rm a b}$}
\end{center}
\vskip2mm
\begin{center}
\footnotesize
\begin{tabular}{ll}
a\ &Department of Physics, Brunel University, Uxbridge, Middlesex UB8 
      3PH, UK, Email: Kamrul.Hassan@brunel.ac.uk \\
b\ &Department of Physics,
Shahjalal Science and Technology University, Sylhet, Bangladesh
\end{tabular}
\end{center}
\vskip40mm
\centerline {\bf {Abstract}} 
\begin{quote}
{\small\rm\  We present two classes of exact solutions to a geometric 
model which describes the kinetics of fragmentation of $d$-dimensional 
hypercuboid-shaped objects. The first class of exact solutions is described by a 
fragmentation rate $a({x_1},...,{x_d}) = 1$ and daughter distribution function 
$b({x_1},..,{x_d} | {{x_{1}^{\p}}},...,{{x_{d}^{\p}}})= 
{{(\a_1 + 2)x_1^{\a_1}}\over{x_1^{\p(\a_1+1)}}}...{{(\a_d+2)x_d^{\a_d}}\over{x_d^{\p(\a_d+1)}}}$.
The second class of exact solutions is described by a 
fragmentation rate $ a({x_1},...,{x_d}) = {{{x_1}^{\a_1}}...{{x_d}^{\a_d}}/{2^d}}$ 
and a daughter distribution function $b({x_1},..,{x_d} | 
{{x_{1}^{\p}}},...,{{x_{d}^{\p}}}) = {2^d}{\d(x_1 - 
{{x_{1}^\p}}/2)...\d(x_d - {{x_{d}^\p}}/2)}$. Each class of exact solutions is 
analyzed in detail for the 
presence of scaling solutions and the occurrence of shattering transitions; the results of these analyses 
are also presented.  }
\end{quote} 
\vskip60mm

PACS: 0520, 0250.

\end{titlepage}

%=======================================================================

\section{Introduction}
\vspace{5mm}

Fragmentation occurs in numerous important physical (droplet 
break-up [1] and fiber length reduction [2]), chemical 
(depolymerization 
through, shear action [3,4], chemical attack [5] and exposure to nuclear, 
ultra-violet and ultra-sonic radiation [6,7]) and geological (rock 
crushing and grinding (communition) [2]) processes. Theoretical 
predictions of the evolution with time of the size distributions of the 
fragmenting objects during such processes is of great interest and 
importance. There are essentially two approaches in use for determining 
the evolution in time of the object size distributions as a function of 
the initial conditions and the fragmentation rates. The first approach 
relies upon statistical and combinatorial arguments [8--10]. The second 
approach has been through the analysis of the kinetic equation modelling 
the fragmentation [11--13].

In the kinetic equation approach the fragmentation process 
can be described by the evolution in time of the size distribution 
$c(x,t)$, where $x$ is the size of the fragments and $t$ is the time, 
through a kinetic equation. This theoretical approach is of a mean field 
nature since fluctuations are ignored. Fragments are assumed to be 
distributed homogeneously at all times throughout the system, i.e., there 
is perfect mixing and the shape of the fragments is ignored. 
Consequently, the size of the fragments is the only dynamical variable 
that characterizes a fragment in the kinetic equation approach. A number 
of authors have expended much effort in finding exact solutions to the 
kinetic equation, in order to study specific practical problems and to 
provide a greater understanding of the behaviour of physical, chemical 
and geological systems in which fragmentation occurs [5,14--21]. Although 
the basic kinetic equations are linear, and in principle soluble, the 
number of exact solutions is few, mainly because of the non-local 
structure of the kinetic equations.

Of considerable importance are the scaling solutions. These 
are essentially the solutions in the long-time ($t \rightarrow\infty$), 
small-size ($x \rightarrow 0$) limit where the size distribution $c(x,t)$ 
evolves to a simpler universal form. This form is universal in the sense 
that it does not depend on the initial conditions. Most experimental 
systems evolve to the point where this behaviour is reached. Scaling 
theories based on a linear kinetic equation have been derived for a large 
class of models which undergo fragmentation [18, 22--24].

The time evolution of the fragmentation process depends 
qualitatively on the behaviour of the probability of the break-up for the 
fragments. For break-up rates increasing sufficiently quickly with 
decreasing size or mass, a cascading break-up occurs in which a finite 
part of the total size or mass is transfered to fragments of zero or 
infinitesimal size or mass. This so-called ``shattering'' [17, 25] or 
``disintegration'' [26] phenomenon is accompanied by a violation of the 
usual dynamical scaling as well as violation of size or mass conservation.

The general form of the 1-dimensional multiple fragmentation 
equation is given by

\be
 {{\partial c(x,t)}\over{\partial t}} = -a(x)c(x,t) + 
         \int_{x}^{\infty} {\rm d}x^{\p}
             a(x^{\p}) b(x|x^{\p}) c(x^{\p},t)
\ee 
where $a(x)$ gives the rate of fragmentation of particles of size $x$, 
the daughter distribution function $b(x|x^{\p})$ is the average 
number of particles of size $x$ produced when a particle of size 
$x^{\p}$ breaks up, and $c(x,t)$ is the size distribution of particles 
of size $x$ at time $t$. To ensure that the total size or mass of the 
fragmenting particles is conserved we insist that 

\be
    x = \int_{0}^{x}{\rm d} x^{\p} x^{\p} b(x^{\p}|x)
\ee
holds. The average number of particles produced when a particle of size 
$x$ fragments is given by

\be
  <N(x)> = \int_{0}^{x} {\rm d} x^{\p} b(x^{\p}|x).
\ee                          
On physical grounds we must have $<N(x)> \geq 2$, which together 
with (2) and (3), places constraints on the possible choices for 
$b(x|x^{\p})$ available to us.

In the special case of binary fragmentation where 2 
particles are produced per fragmentation event (1--3) can be 
rewritten in terms of the single symmetric function $F(x,x^{\p}) 
= F(x^{\p},x)$ as follows. Firstly, write, 

\be
  a(x) = \int_{0}^{x}{\rm d}{x^{\p}}F(x-x^{\p},x^{\p}).
\ee           
Then to ensure that there are precisely 2 particles produced per 
fragmentation event, choose

\be
  b(x|x^{\p}) = {{2F(x,x^{\p}-x)}\over{a(x^{\p})}},
\ee 
in which case (1) becomes

\be
{{\partial c(x,t)}\over{\partial t}} = -c(x,t) \int_{0}^{x}{\rm 
     d}{x^{\p}}F(x-x^{\p},x^{\p}) + 2\int_{x}^{\infty}{\rm d}x^{\p}
          F(x,x^{\p}-x)c(x^{\p},t)
\ee
where $F(x,x^{\p})$ describes the rate at which particles of size 
$(x+x^{\p})$ fragment into particles of size $x$ and $x^{\p}$.

As mentioned above, the kinetics of such 1-dimensional 
fragmentation processes is now well-understood with numerous explicit 
exact solutions, scaling solutions and quantitative descriptions of 
shattering transitions known to us.

In realistic fragmentation processes particles have both size 
and shape, and it is clear that this geometry of these fragmenting 
particles will influence the fragmentation process. For example a particle 
may be selected for fragmentation at a rate which is dependent on its 
area or volume, but the manner in which the fragmentation of the particle 
is implemented will, in general, depend on its precise dimensions. If 2 
particles have the same area but one is needle-shaped and the other is 
square-shaped, they may be equally likely to fragment as a consequence of 
their possesing the same area, but the needle-shaped particle is much 
more likely to fragment across its longer side, whereas the square-shaped 
particle is equally likely to fragment across either side. Until 
recently, all these properties were represented by a single parameter, 
namely the size or mass of the fragmenting particle.

Recently, various authors [27--29] have introduced and 
investigated simple kinetic models describing the fragmentation of 
2-dimensional and more generally $d$-dimensional particles. These authors 
present several simple classes of explicit exact solutions for 
2-dimensional models. Krapivsky and Ben-Naim [28] also discuss the 
presence of scaling and multi-scaling in their models of fragmentation 
for $d$-dimensional particles. In [29], the shattering transition in a 
2-variable fragmentation model is investigated.

However, as far as we know there do not exist any explicit 
exact solutions for fragmenting particles in $d$-dimensions, for general 
$d$. We present in this paper, 2 classes of exact solutions to a 
geometric model which describes the kinetics of fragmentation of 
$d$-dimensional objects, for general $d$. Each class of exact solutions 
is analyzed in detail for the presence of scaling, multi-scaling,
and the occurrence of shattering transitions.

%=======================================================================

\section{Fragmentation in $d$-dimensions}
\vspace{5mm}
The general form of the $d$-dimensional multiple 
fragmentation equation is given by

\begin{eqnarray}
{{\partial c(x_1,...,x_d,t)}\over{\partial t}} & = & 
      - a(x_1,...,x_d)c(x_1,...,x_d,t) \nonumber  \\  
   & + &  \int_{x_1}^{\infty} {\rm d}x_1^{\p}...
         \int_{x_d}^{\infty}{\rm d}x_d^{\p}a(x_1^{\p},...,x_d^{\p})
              b(x_1,...,x_d|x_1^{\p},...,x_d^{\p})
                   c(x_1^{\p},...,x_d^{\p},t)
\end{eqnarray}
where $c(x_1,...,x_d,t)$ is the size distribution of particles with size 
$x_1...x_d$ at time $t$, $a(x_1,...,x_d)$ is the rate at which a particle 
characterized by $(x_1,...,x_d)$ fragments, and the daughter distribution 
function $b(x_1,...,x_d|x_1^{\p},...,x_d^{\p})$ is the rate at which a 
particle characterized by $(x_1,...,x_d)$ is produced from a particle 
characterized by $(x_1^{\p},...,x_d^{\p})$.

Consequently, the average number of particles produced per 
fragmentation event is

\be
 <N(x_1,...,x_d)> = \int_{0}^{x_1} {\rm d} x_1^{\p}...
      \int_{0}^{x_d}{\rm d}x_d^{\p}
          b(x_1^{\p},...,x_d^{\p}|x_1,...,x_d)
\ee
and mass conservation requires,

\be
  x_1...x_d = \int_{0}^{x_1} {\rm d} x_1^{\p}...
      \int_{0}^{x_d}{\rm d}x_d^{\p}x_1^{\p}...x_d^{\p}
          b(x_1^{\p},...,x_d^{\p}|x_1,...,x_d).
\ee
Equations (7--9) form a complete set of equations, which define the 
fragmentation process given $a(x_1,...,x_d)$, 
$b(x_1,...,x_d|x_1^{\p},...,x_d^{\p})$ and suitable initial conditions. 
Of course, the functions $a(x_1,...,x_d)$ 
and$b(x_1,...,x_d|x_1^{\p},...,x_d^{\p})$ must be chosen to make the 
equations physically meaningful and furthermore 
$b(x_1,...,x_d|x_1^{\p},...,x_d^{\p})$ must be chosen to ensure that 
$(8)$ and $(9)$ hold, as well as the obvious physical constraint 
$N(x_1,...,x_d) \geq 2$.

As a special case, consider fragmentation processes such that 
a given $d$-dimensional particle fragments into $2^d$ pieces per 
fragmentation event. In this case, as in section 1 above, we may rewrite 
(7--9) in terms of a single function $F(x_1,x_1^{\p};..;x_d,x_d^{\p})$ as 
follows. Firstly, choose

\be
 a(x_1,...,x_d) = 
         \int_{0}^{x_1} {\rm d} x_1^{\p}...
            \int_{0}^{x_d}{\rm d}x_d^{\p}
               F(x_1-x_1^{\p},x_1^{\p};..;x_d-x_d^{\p},x_d^{\p})
\ee
where $F(x_1,x_1^{\p};..;x_d,x_d^{\p})$ is the rate of fragmentation of 
a particle characterized by $(x_1+x_1^{\p})...(x_d+x_d^{\p})$ into $2^d$ 
smaller particles characterized by: $x_1...x_d$,..., 
$x_1^{\p}...x_d^{\p}$. Now choose,

\be
  b(x_1,...,x_d|x_1^{\p},...,x_d^{\p}) = 
        {{2^d}{F(x_1,x_1^{\p}-x_1;..;x_d,x_d^{\p}-x_d)}
             \over{a(x_1^{\p},...,x_d^{\p})}}
\ee
where, of course,

\be
 F(x_1,x_1^{\p}-x_1;..;x_k,x_k^{\p};...;x_d,x_d^{\p}-x_d) =
              F(x_1,x_1^{\p}-x_1;..;x_k^{\p},x_k;...;x_d,x_d^{\p}-x_d)
\ee
i.e., $F(x_1,x_1^{\p}-x_1;..;x_k,x_k^{\p};...;x_d,x_d^{\p}-x_d)$ is 
symmetric in all pairs of arguments $(x_k,x_k^{\p})$ for $k = 1,...,d$.

It is easily shown that $<N(x_1,...,x_d)> = 2^d$, as 
required, and that mass conservation requires,

\be
  x_1...x_d = {{{2^d}{ \int_{0}^{x_1} {\rm d} x_1^{\p}...
            \int_{0}^{x_d}{\rm d}x_d^{\p}x_1^{\p}...x_d^{\p}
               F(x_1-x_1^{\p},x_1^{\p};..;x_d-x_d^{\p},x_d^{\p})}}
                 \over{\int_{0}^{x_1} {\rm d} x_1^{\p}...
            \int_{0}^{x_d}{\rm d}x_d^{\p}
               F(x_1-x_1^{\p},x_1^{\p};..;x_d-x_d^{\p},x_d^{\p})}}.
\ee

The fragmentation equation (7) now becomes,

\begin{eqnarray}
{{\partial c(x_1,...,x_d,t)}\over{\partial t}} & = &
   - c(x_1,...,x_d,t)\int_{0}^{x_1} {\rm d} x_1^{\p}...
      \int_{0}^{x_d}{\rm d}x_d^{\p}
          F(x_1-x_1^{\p},x_1^{\p};..;x_d-x_d^{\p},x_d^{\p}) \nonumber  \\
   & + & {2^d} \int_{x_1}^{\infty} {\rm d}x_1^{\p}...
         \int_{x_d}^{\infty}{\rm d}x_d^{\p}
             F(x_1^{\p}-x_1,x_1;...;x_d^{\p}-x_d,x_d)
                   c(x_1^{\p},...,x_d^{\p},t)
\end{eqnarray}
where $F(x_1,x_1^{\p};..;x_d,x_d^{\p})$ is defined by (10--13).

\vspace{5mm}

\begin{flushleft}
{\large\bf Model 1}
\vskip3mm
\end{flushleft}
This model is described by the fragmentation rate and 
daughter distribution function given, respectively, by

\begin{eqnarray}
                            a(x_1,...,x_d) & = & 1    \\
      b(x_1,...,x_d|x_1^{\p},...,x_d^{\p}) & = &
            {(\a_1+2)...(\a_d+2){x_1^{\a_1}...x_d^{\a_d}}
               \over{x_1^{\p\a_1+1}...x_d^{\p\a_d+1}}}.      
\end{eqnarray}
Insisting that mass is conserved per fragmentation event places the 
following restrictions on the homogeneity indices $\a_1,...,\a_d$, 
\be
 \a_i > -2  \hspace{35mm} ;i = 1,...,d.
\ee
It is easily shown that the number of fragments produced per 
fragmentation event is
 
\be              
 <N(x_1,...,x_d)> = \left\{ \begin{array}{r@{\qquad;\qquad}l}
   {{(\a_1+2)...(\a_d+2)}\over{(\a_1+1)...(\a_d+1)}}
        &  \mbox{if all }  \a_i > -1   \\
     \infty\hfill &  \mbox{if some or all }   \a_i \leq -1 \end{array} 
\right. . 
\ee
On physical grounds, for all $\a_i > -1$, we must insist that when 
$<N(x_1,...,x_d)>$ is finite it must be so that $<N(x_1,...,x_d)> \geq 
2$. This constrains the $\a_i$ by

\be  
{{(\a_1+2)...(\a_d+2)}\over{(\a_1+1)...(\a_d+1)}} \geq 2.
\ee
The $d$-dimensional multiple fragmentation equation (7) now becomes

\begin{eqnarray}
{{\partial c(x_1,...,x_d,t)}\over{\partial t}} & & =
   - c(x_1,...,x_d,t) \nonumber  \\
     + & & (\a_1+2)...(\a_d+2)x_1^{\a_1}...x_d^{\a_d} 
       \int_{x_1}^{\infty}{{\rm d}x_1^{\p}\over{x_1^{\p\a_1+1}}}...
         \int_{x_d}^{\infty}{{\rm d}x_d^{\p}\over{x_d^{\p\a_d+1}}}
             c(x_1^{\p},...,x_d^{\p},t).
\end{eqnarray}
We need to solve (20) subject to appropriate initial conditions,

\be
   c(x_1,...,x_d,0) = f(x_1,...,x_d) \not= 0.
\ee

Define the Laplace transform of $c(x_1,...,x_d,t)$ with respect to $t$,
$\phi(x_1,...,x_d,s)$, by

\be
  \phi(x_1,...,x_d,s) = \int_{0}^{\infty}{\rm d}t e^{-st} c(x_1,...,x_d,t)
\ee
in which case we may recover $c(x_1,...,x_d,t)$ from the inverse Laplace 
transform

\be
  c(x_1,...,x_d,t) = {1\over{2\pi i}}
      \int_{\gamma - i\infty}^{\gamma + i\infty}
         {\rm d}s e^{st}\phi(x_1,...,x_d,s)
\ee
where $\Re(s) > \gamma $ to ensure convergence.

Taking the Laplace transform of (20) with respect to $t$ yields

\begin{eqnarray}
\phi(x_1,...,x_d,s) & & =
    {f(x_1,...,x_d)\over{(s+1)}} \nonumber  \\
     + & & {(\a_1+2)...(\a_d+2)x_1^{\a_1}...x_d^{\a_d}\over{(s+1)}}
       \int_{x_1}^{\infty}{{\rm d}x_1^{\p}\over{x_1^{\p\a_1+1}}}...
         \int_{x_d}^{\infty}{{\rm d}x_d^{\p}\over{x_d^{\p\a_d+1}}}
             \phi(x_1^{\p},...,x_d^{\p},s).
\end{eqnarray}
Following the approach introduced in [30] one finds

\begin{eqnarray}
\phi(x_1,...,x_d,s) & & =
    {f(x_1,...,x_d)\over{(s+1)}} \nonumber  \\
     + & & {(\a_1+2)...(\a_d+2)x_1^{\a_1}...x_d^{\a_d}\over{{(s+1)}^2}}
       \int_{x_1}^{\infty}{{\rm d}x_1^{\p}\over{x_1^{\p\a_1+1}}}...
         \int_{x_d}^{\infty}{{\rm d}x_d^{\p}\over{x_d^{\p\a_d+1}}}
             f(x_1^{\p},...,x_d^{\p}) \nonumber \\
     \times & & \sum_{r=0}^{\infty}
               {1\over{{(r!)}^d}}{\Bigg [
                 {{(\a_1+2)...(\a_d+2)\over{(s+1)}}}
                   {\ln \Big ({x_1^{\p}\over{x_1}}\Big )}...
                     {\ln \Big ( {x_d^{\p}\over{x_d}}\Big )} \Bigg ]}^r . 
\end{eqnarray}
Performing a simple contour integration yields
 
\begin{eqnarray}
  c(x_1,...,x_d,t) & & =
   e^{-t} \Bigg (f(x_1,...,x_d)
    +(\a_1+2)...(\a_d+2)x_1^{\a_1}...x_d^{\a_d} \nonumber \\
     \times & & \int_{x_1}^{\infty}{{\rm d}x_1^{\p}\over{x_1^{\p\a_1+1}}}...
        \int_{x_d}^{\infty}{{\rm d}x_d^{\p}\over{x_d^{\p\a_d+1}}}
          f(x_1^{\p},...,x_d^{\p}) \nonumber \\
        \times & &  \sum_{r=0}^{\infty}
           {t^{r+1}\over{(r+1)!{(r!)}^d}}{\Bigg [
             {(\a_1+2)...(\a_d+2)}
                {\ln \Big ({x_1^{\p}\over{x_1}}\Big )}...
                  {\ln \Big ( {x_d^{\p}\over{x_d}}\Big )} \Bigg ]}^r \Bigg) .
\end{eqnarray}
For mono-disperse initial conditions,

\be
 f(x_1,...,x_d) = \d (x_1-l_1)...\d (x_d-l_d)
\ee
(26) becomes

\begin{eqnarray}
  c(x_1,...,x_d,t) & & =
   e^{-t} \Bigg (\d (x_1-l_1)...\d (x_d-l_d)
    +(\a_1+2)...(\a_d+2)
  {x_1^{\a_1}...x_d^{\a_d}\over{l_1^{\a_1+1}...l_d^{\a_d+1}}} \nonumber \\
        \times & &  \sum_{r=0}^{\infty}
           {t^{r+1}\over{(r+1)!{(r!)}^d}}{\Bigg [
             {(\a_1+2)...(\a_d+2)}
                {\ln \Big ({l_1\over{x_1}}\Big )}...
                  {\ln \Big ( {l_d\over{x_d}}\Big )} \Bigg ]}^r \Bigg) .
\end{eqnarray}
For $\a_i = 0$ and $d = 1$ (28) reduces to the result of Ziff and McGrady 
[16]. For $\a_i = 0$ and $d = 2$ (28) reduces to the solution presented by 
Rodgers and Hassan [27]. When the $\a_i$ are completely general and $d = 1$,
(28) is equivalent to the exact solution for the model investigated by 
McGrady and Ziff [17], with $\beta = -1$.

\vspace{5mm}

\begin{flushleft}
{\large\bf Model 2}
\vskip3mm
\end{flushleft}

In this model we investigate a fragmentation rate and daughter 
distribution function given, respectively, by

\begin{eqnarray}
         a(x_1,...,x_d) & = & {1\over{2^d}}x_1^{\a_1}...x_d^{\a_d}    \\
      b(x_1,...,x_d|x_1^{\p},...,x_d^{\p}) & = &
            2^d{\d (x_1-{{x_1^{\p}}\over{2}})...\d 
                                     (x_d-{{x_d^{\p}}\over{2}})}.
\end{eqnarray}
It is easily shown that mass conservation per single fragmentation event 
holds, and that the average number of particles per fragmentation event,
$<N(x_1,...,x_d)>$, is $2^d$.

This particular choice for the fragmentation rate $a(x_1,...,x_d)$ and 
the daughter distribution function $b(x_1,...,x_d | 
x_1^{\p},...,x_d^{\p})$ can be implemented by 
$F(x_1,x_1^{\p};...;x_d,x_d^{\p})$ with

\be
 F(x_1,x_1^{\p};...;x_d,x_d^{\p}) = 
        {(x_1+x_1^{\p})}^{\a_1}...{(x_d+x_d^{\p})}^{\a_d}
          {\d (x_1-x_1^{\p})}...{\d (x_d-x_d^{\p})}.
\ee
In this form the kinetics of the model become a little more transparent. 
The fragmentation rate $F(x_1,x_1^{\p};...;x_d,x_d^{\p})$ describe a 
fragmentation process in which an object splits into $2^d$ equally(?) 
sized pieces. 

This choice for $a(x_1,...,x_d)$ and $b(x_1,...,x_d |
x_1^{\p},...,x_d^{\p})$, or equivalently, $F(x_1,x_1^{\p};...;x_d,x_d^{\p})$,
reduces the $d$-dimensional multiple fragmentation equation (7) to the 
following form,

\be
  {{\partial c(x_1,...,x_d,t)}\over{\partial t}} = 
      -{{{x_1^{\a_1}}...{x_d^{\a_d}}}\over{2^d}}c(x_1,...,x_d,t) +
       2^{d+\a} {{x_1^{\a_1}}...{x_d^{\a_d}}}c(2x_1,...,2x_d,t)
\ee
where $\a = \a_1 +...+\a_d$.
 
Solving (32) subject to the initial conditions

\be
  c(x_1,...,x_d,0) = f(x_1,...,x_d) \not= 0
\ee
via the approach outlined in [30], one finds

\begin{eqnarray}
  c(x_1,...,x_d,t) &=& e^{-{{{x_1^{\a_1}}...{x_d^{\a_d}}t}/{2^d}}} 
        \Bigg (f(x_1,...,x_d) \nonumber \\
     &+& \sum_{r=1}^{\infty} 2^{{{r(r+1)\a}/{2}}+2rd}
                    f(2^r x_1,...,2^r x_d)
         \sum_{k=0}^{r}
          {{e^{{{x_1^{\a_1}}...{x_d^{\a_d}}(1-2^{k\a})t}/{2^d}}}
         \over{\P (2^{m \a}-2^{k \a})}} \Bigg )
\end{eqnarray}
with $\a = \a_1 +...+\a_d \not= 0$ and $ I_{k}^{r} = 
\{0,1,2,...,k-1,k+1,...,r\}$.

Using the fact that

\be
 \lim_{\a \to 0} \sum_{k=0}^{r}
          {{e^{{{x_1^{\a_1}}...{x_d^{\a_d}}(1-2^{k\a})t}/{2^d}}}
         \over{\P (2^{m \a}-2^{k \a})}} = \Bigg ( 
            {{{x_1^{\a_1}}...{x_d^{\a_d}}t}\over{2^d}} \Bigg ) ^r 
                {1\over{r!}}
\ee
we can obtain $c(x_1,...,x_d,t)$ for $\a = \a_1 +...+ \a_d = 0$ without 
any extra effort. Explicitly,

\be
  c(x_1,...,x_d,t) = e^{-{{{x_1^{\a_1}}...{x_d^{\a_d}}t}/{2^d}}}
     \sum_{r=0}^{\infty}{{(2^d t)^r}\over{r!}}
          {({x_1^{\a_1}}...{x_d^{\a_d}})}^r
             f(2^r x_1,...,2^r x_d)
\ee
where, of course, $\a = \a_1 +...+ \a_d$.

For mono-disperse initial conditions,

\be
  f(x_1,...,x_d) = \d (x_1-l_1)...\d (x_d-l_d)
\ee
we find 

\be
  c(x_1,...,x_d,t) = \left\{ \begin{array}{r@{\quad\quad}l}
       e^{-{{{l_1^{\a_1}}...{l_d^{\a_d}}t}/{2^d}}}
          \Bigg (\d (x_1-l_1)...\d (x_d-l_d) + \sum\limits_{r=1}^{\infty} 
             2^{{{r(r+1)\a}/{2}}+rd}   & \\
              \times \d (x_1-{{l_1}\over{2^r}})...\d (x_d-{{l_d}\over{2^r}}) 
           \sum\limits_{k=0}^{r}
          {{e^{{{l_1^{\a_1}}...{l_d^{\a_d}}(1-2^{(k-r)\a})t}/{2^d}}}
         \over{\P (2^{m \a}-2^{k \a})}} \Bigg ) &  {\rm ;} \a \not= 0   \\
     e^{-{{{l_1^{\a_1}}...{l_d^{\a_d}}t}/{2^d}}}
        \sum\limits_{r=0}^{\infty}{t^r\over{r!}}
         ({l_1^{\a_1}}...{l_d^{\a_d}})^r
             \d (x_1-{{l_1}\over{2^r}})...\d (x_d-{{l_d}\over{2^r}})
                \hfill & {\rm ;}  \a = 0 \end{array} \right..
\ee
For $d = 1$, these results reduce to those presented in [30]. For $d = 1$
and $\a = \a_1 = 0$, one recovers the exact solution of Bak and Bak [19].

%=======================================================================

\section{Scaling and multi-scaling}
\vspace{5mm}

To investigate the presence of scaling or multi-scaling, we introduce 
the $d$-tuple Mellin transform of the distribution function 
$c(x_1,...,x_d,t)$ defined by

\be
 M(s_1,...,s_d,t) = \int_{0}^{\infty}{\rm d}x_1...
       \int_{0}^{\infty}{\rm d}x_d {x_1^{s_1-1}}...{x_d^{s_d-1}}
            c(x_1,...,x_d,t).
\ee
The functions $M(s_1,...,s_d,t)$ for fixed $s_1,...,s_d$ are known as the 
moments of the distribution function $c(x_1,...,x_d,t)$.

Combining (7) and (39) gives

\begin{eqnarray}
  {{\partial M(s_1,...,s_d,t)}\over{\partial t}}  &=&  - 
    \int_{0}^{\infty}{\rm d}x_1...\int_{0}^{\infty}{\rm d}x_d 
     a(x_1,...,x_d)c(x_1,...,x_d,t) \nonumber \\
 \times  \Big ({x_1^{s_1-1}}...{x_d^{s_d-1}} &-&
  \int_{0}^{x_1}{\rm d}x_1^{\p}...\int_{0}^{x_d}{\rm d}x_d^{\p}
   {x_1^{\p{s_1-1}}}...{x_d^{\p{s_d-1}}}
b(x_1^{\p},...,x_d^{\p} | x_1,...,x_d) \Big ).
\end{eqnarray}
We now investigate the the two classes of exact solutions to the 
$d$-dimensional multiple fragmentation equation, presented in section 2, 
for the presence of scaling and multi-scaling.

\vspace{5mm}

\begin{flushleft}
{\large\bf Model 1}
\vskip3mm
\end{flushleft} 

In this case (40) becomes,

\be
 {{\partial M(s_1,...,s_d,t)}\over{\partial t}} = 
    -\Bigg (1-{{(\a_1+2)...(\a_d+2)}\over{(s_1+\a_1)...(s_d+\a_d)}}\Bigg )
       M(s_1,...,s_d,t)
\ee
provided $s_i+\a_i > 0$ for $i = 1,...,d$. Then it follows that

\be
  M(s_1,...,s_d,t) = M(s_1,...,s_d,0) e^{-\big 
    (1-{{(\a_1+2)...(\a_d+2)}\over{(s_1+\a_1)...(s_d+\a_d)}}\big )t}
\ee
provided $s_i+\a_i > 0$ for $i = 1,...,d$.

To obtain the number of objects,$N(t)$, in our fragmenting system we must
take $s_i = 1$ for $i =1,...,d$ in (42). We find that

\be
  N(t) = M(1,...,1,t) = M(1,...,1,0)e^{-\big
    (1-{{(\a_1+2)...(\a_d+2)}\over{(\a_1+1)...(\a_d+1)}}\big )t}
\ee
which is only valid for $\a_i > -1$ and $i = 1,...,d$. When some or all 
of the $\a_i \leq -1$, then (41) and (42) are not valid, since it can 
easily be shown that the number of particles in the system is infinite,
even though the mass or volume, $V(t)$, of the system is conserved. If we 
set $s_i = 2$ for $i = 1,...,d$ in (42) we find that the volume, $V(t)$,
of our system is constant, i.e.,

\be
  V(t) = M(2,...,2,t) = M(2,...,2,0)
\ee
provided $\a_i > -1$ for $i = 1,...,d$.

An interesting feature of (41), for $s_i+\a_i > 0$ and $i = 1,...,d$, is 
that it implies the existence of an infinite number of conservation laws, 
apart from the usual volume conservation one usually encountered. The 
moments $M(s_1,...,s_d,t)$ with $s_1,...,s_d$ satisfying

\be
  {{(\a_1+2)...(\a_d+2)}\over{(s_1+\a_1)...(s_d+\a_d)}} = 1
\ee
are all time-independent. Of course, $s_i \not= 1$ $\forall i = 1,...,d$ in 
this case, otherwise (45) would contradict (19) even if $\a_i > -1$ 
$\forall i = 1,...,d$. Besides, the number of objects, $N(t)$, in a 
fragmenting system cannot possibly be conserved. Thus, in addition to the 
conservation of the total volume, $V(t)$, there are an infinite number of 
hidden conserved integrals for those $s_i$, with $i = 1,...,d$, that lie 
on the hypersurface defined by (45). According to Krapivsky and Ben-Naim 
[28], it is precisely these integrals which are responsible for the 
absence of scaling solutions for the $d$-dimensional fragmentation 
processes. Indeed, the scaling solution

\be
  c(x_1,...,x_d,t) \sim t^w {\phi(t^zx_1,...,t^zx_d)}
\ee 
implies an infinite number of scaling relations

\be
  w = zs
\ee
where $s = s_1 + ... + s_d$, which together with (45) cannot all be 
satisfied by the scaling exponents $w$ and $z$. This rules out the 
possibility of scaling solutions in this model, however multi-scaling 
solutions may be possible.

\vspace{5mm}

\begin{flushleft}
{\large\bf Model 2}
\vskip3mm
\end{flushleft}

For this case (40) becomes,

\be
  {{\partial M(s_1,...,s_d,t)}\over{\partial t}} =
    \Bigg ({1\over{2^{(s-d)}}}-{1\over{2^d}}\Bigg )
       M(s_1+\a_1,...,s_d+\a_d,t)
\ee
where $s = s_1+...+s_d$. Again, as in model 1, we observe that (48) 
implies the existence of an infinite number of conservation laws. The 
moments $M(s_1,...,s_d,t)$ with the $s_i$, for $i = 1,...,d$, satisfying

\be
 s = 2d
\ee 
where $s = s_1+...+s_d$, are all time-independent. Thus, in addition to 
the conservation of the total volume, $V(t) = M(2,...,2,t)$, there are an 
infinite number of hidden conserved integrals for those $s_i$, with $ i = 
1,...,d$, that lie on the hypersurface defined by (49). Due to the 
existence of an infinite number of hidden conserved integrals, we do not 
expect there to be any scaling solutions to this model. However, this 
does not appear to be the case, and we demonstrate this explicitly by 
finding a scaling solution to this model.

In this case assume a scaling solution of the form

\be
  c(x_1,...,x_d,t) \sim {V(t)}^{-2d} \phi (x_1/V(t),...,x_d/V(t))
\ee 
as $t\rightarrow \infty$, where $V(t)$ is the volume of a typical 
time-dependent cluster volume. This scaling form is fully equivalent to (46),
however its form is more convenient for our investigations [18]. The 
exponent $-2d$ ensures conservation of volume of the complete system of 
fragmenting particles.

A short calculation will show that 

\be
  M(s_1,...,s_d,t) \sim {V(t)}^{s-2d} m(s_1,...,s_d)
\ee
where the scaling moments $m(s_1,...,s_d)$ are defined by

\be
 m(s_1,...,s_d) = \int_{0}^{\infty}{\rm d}\xi_1 ...
       \int_{0}^{\infty}{\rm d}\xi_d {\xi_1}^{s_1-1}...{\xi_d}^{s_d-1}
             \phi(\xi_1,...,\xi_d).
\ee

If the moments are to be conserved, then they must be constant. This 
occurs when

\be
  s = 2d
\ee
which is in complete agreement with (49), as it should be.

Substituting (50) into (32) yields

\begin{eqnarray}
 -{1\over{{V(t)}^{\a+1}}}{{\rm d}V(t)\over{{\rm d}t}} &=& \omega 
\nonumber \\ 
   & =& {{\Big ( -{1\over{2^d}}\phi (\xi_1,...,\xi_d) +{2^{d+\a}}\phi 
    (2\xi_1,...,2\xi_d) \Big ) \xi_1^{\a_1}...\xi_d^{\a_d}}
      \over{\Big (\xi_i {{\partial \phi (\xi_1,...,\xi_d)}\over{\partial 
        \xi_i}} + 2d \phi (\xi_1,...,xi_d) \Big )}} 
\end{eqnarray}
where the separation constant $\omega$ is positive since $V(t)$ must be a 
decreasing function of time in a fragmenting system and $\xi_i = x_i/V(t)$
for $i = 1,...,d$. Then

\be
  V(t) \sim  \left\{ \begin{array}{r@{\qquad\qquad}l}
      t^{-{1\over{\a}}} \hfill & {\rm ;} \hspace{5mm} \a > 0 
        {\rm ,} \hspace{5mm} t \rightarrow \infty \\
          e^{-\omega t} \hfill &  {\rm ;} \hspace{5mm} \a = 0
            {\rm ,} \hspace{5mm} t \rightarrow \infty   \\
         (t_c-t)^{1\over{|\a|}} & {\rm ;} \hspace{5mm}  \a < 0
               {\rm ,} \hspace{5mm} t < t_c \end{array} \right..
\ee 
These expressions are only valid provided scaling holds. For $\a < 0$ a 
singularity is encountered within a finite time $t_c$, and scaling 
becomes invalid. In this instance, we anticipate shattering which will be 
discussed later. When $\a = 0$, as in the 1-dimensional case [24, 26, 
30], a scaling form does not exist which is consistent with $\phi 
(\xi_1,...,\xi_d) \rightarrow $ constant as $\xi_k \rightarrow 0$ and
$\phi (\xi_1,...,\xi_d) \rightarrow 0$ as $\xi_k \rightarrow \infty$ for 
$k = 1,...,d$. Consequently, we need only look at the case when $\a > 0$, 
for which scaling is valid. 

When $\a > 0$, we assume that $\phi (\xi_1,...,\xi_d)$ vanishes at $\xi_k 
= 0$ and $\xi_k = \infty$ for $k = 1,...,d$. It can be shown from (54) 
that 

\be
  \phi (\xi_1,...,\xi_d) \sim
   {{e^{-\xi_1^{\a_1}...\xi_d^{\a_d}/{2^d \omega \a}}}
     \over{(\xi_1^2 + ... +\xi_d^2)^d}}
\ee
as $\xi_k \rightarrow \infty$ for $ k = 1,...,d$. Therefore, we will 
assume that

\be 
  \phi (\xi_1,...,\xi_d) =
   {{e^{-\xi_1^{\a_1}...\xi_d^{\a_d}/{2^d \omega \a}}}
     \over{(\xi_1^2 + ... +\xi_d^2)^d}}f(\xi_1,...,\xi_d)
\ee
where we insist that $f(\xi_1,...,\xi_d) = 1$ at $\xi_k = \infty$ for $ k 
= 1,...,d$.

Sustituting (57) into (54) and performing a lengthy calculation yields a 
solution for $f(\xi_1,...,\xi_k)$. Hence, for $\a > 0$,

\be
\phi (\xi_1,...,\xi_d) =
   {{e^{-\xi_1^{\a_1}...\xi_d^{\a_d}/{2^d \omega \a}}}
     \over{(\xi_1^2 + ... +\xi_d^2)^d}}
      {\Bigg (1 + \sum\limits_{n=1}^{\infty}
        {(-1)^n 2^{n\a}\over{\PQ(2^{m\a}-1)}}
          {e^{-\xi_1^{\a_1}...\xi_d^{\a_d}(2^{n\a}-1)/{2^d \omega \a}}} 
                   \Bigg ) }.
\ee 
Thus, a rather unusual situation occurs in this $d$-dimensional model. A 
scaling solution exists, in spite of the fact that we have an infinite
number of hidden conserved integrals, of the form (50), as $t \rightarrow 
\infty$ with $V(t)$ given by (55) and $\phi (\xi_1,...,\xi_d)$ given by (58).
These results are consistent with the exact solution of model 2 presented 
in section 2 above. When $d = 1$ these results reduce to those in [30] 
and are of a similar form to those of Cheng and Redner [24].

%=======================================================================

\section{Shattering transitions}
\vspace{5mm}

Formally, the volume, $V(t)$, of the system is defined by

\be
  V(t) = \int_{0}^{\infty}{\rm d}x_1...\int_{0}^{\infty}{\rm d}x_d
          x_1...x_d c(x_1,...,x_d,t),
\ee
so that , with the aid of (7) and (9) it can easily be shown that 

\be
  {{{\rm d}V(t)}\over{{\rm d}t}} = 0
\ee
indicating that the volume, $V(t)$, is conserved. However, when the 
fragmentation rate increases sufficiently fast as the volume of the 
fragments decreases to zero, a cascading of the fragmentation occurs such 
that volume is lost to fragments of zero or inifinitesimal volume. This 
cascading process, which has been named ``shattering'' [17, 25] or 
``disintegration'' [26], is somewhat similar to gelation in coagulating 
systems, where mass is lost to an infinite gel molecule [31, 32]. 
Gelation and shattering are both signalled by the condition ${\rm d}V(t) 
/{\rm d}t < 0$. When shattering is suspected a more subtle analysis to 
that used to derive (60) is required.

To analyze the shattering transition define a cut-off volume 
$V_{\v}(t)$, with $0 < \v \ll 1$, by

\be
  V_{\v}(t) = \int_{\v}^{\infty}{\rm d}x_1...\int_{\v}^{\infty}
      {\rm d}x_d x_1...x_d c(x_1,...,x_d,t),
\ee
with 

\be
  V(t) = \lim_{\v \to 0^{+}} V_{\v}(t).
\ee
It is easily shown that the cut-off volume loss is given by

\begin{eqnarray}
  {{{\rm d}V_{\v}(t)}\over{{\rm d} t}} = 
   &-&\int_{\v}^{\infty}{\rm d}x_1...\int_{\v}^{\infty}
      {\rm d}x_d a(x_1,...,x_d) c(x_1,...,x_d,t) \nonumber \\
      & \times &  \int_{0}^{\v}{\rm d}x_1^{\p}...\int_{0}^{\v}
          {\rm d}x_d^{\p} x_1^{\p}...x_d^{\p}
            b(x_1^{\p},...,x_d^{\p} | x_1,...,x_d)
\end{eqnarray}
where, of course,

\be
 {{{\rm d}V(t)}\over{{\rm d} t}} = \lim_{\v \to 0^{+}}
 {{{\rm d}V_{\v}(t)}\over{{\rm d} t}}.
\ee

\vspace{5mm}

\begin{flushleft}
{\large\bf Model 1}
\vskip3mm
\end{flushleft}

In this model shattering does not occur for any values of the homogeneity 
indices $\a_i$, for $i = 1,...,d$. This is easily demonstrated from the 
definition 
of volume, $V(t)$, of our system via (59) and the exact solution for this 
particular model (26). One finds that the volume, $V(t)$, of our system 
is both finite and time-independent.

\vspace{5mm}

\begin{flushleft}
{\large\bf Model 2}
\vskip3mm
\end{flushleft}

As can be easily demonstrated by substituting the exact solution for this 
particular model (36) into the definition of the volume, $V(t)$, of the 
system (59), shattering does not occur when the sum of the homogeneity 
indices, $\a$, is zero. Again, one finds that the volume, $V(t)$, of the 
system is both finite and time-independent. It remains to investigate the 
case when the sum of the homogeneity indices, $\a$, is less than zero.

To analze the shattering regime, $\a < 0$, we need to know the behaviour 
of $c(x_1,...,x_d,t)$ as $t \rightarrow \infty$ and $x_k \rightarrow 0$
such that $tx_k$ remains fixed for $k = 1,...,d$. A short calculation 
will show that the asymptotic form of $c(x_1,...,x_d,t)$ in the 
shattering regime is

\be
  c(x_1,...,x_d,t) \sim T(t) x_1^{\lambda_1}...x_d^{\lambda_d}
\ee
where $T(t) \not= 0$. The exponents $\lambda_1,...,\lambda_d$ are 
restricted by

\be
  \lambda = |\a| -2d
\ee
where $\lambda = \lambda_1 +...+\lambda_d$. In this analysis it does not 
matter what each individual value of the $\lambda_k$, for $k = 1,...,d$, 
is, all that matters is what the sum, $\lambda$, is. Therefore, we need 
not concern ourselves with explicitly determining the $\lambda_k$, for
$k = 1,...,d$.

For $\a < 0$, (63) becomes 

\be
  {{{\rm d}V_{\v}(t)}\over{{\rm d} t}} =-{1\over{2^d}}
    \int_{\v}^{2\v}{\rm d}x_1...\int_{\v}^{2\v} {\rm d}x_d
       x_1^{\a_1+1}...x_d^{\a_d+1} c(x_1,...,x_d,t) 
\ee
for this particular model. Substituting (65) into (67) yields

\be
  {{{\rm d}V_{\v}(t)}\over{{\rm d} t}} =
                   -{1\over{2^d}}T(t) (\ln 2)^d
\ee
$\forall$ $\a_i + \lambda_i = -2$, with $1 \leq i \leq d$ and

\be
  {{{\rm d}V_{\v}(t)}\over{{\rm d} t}} =
  -{1\over{2^d}}T(t) (\ln 2)^k
          \Big ({2^{(\a_{k+1}+\lambda_{k+1}+2)}-1
            \over{\a_{k+1}+\lambda_{k+1}+2}}\Big )...
             \Big ({2^{(\a_{d}+\lambda_{d}+2)}-1
            \over{\a_{d}+\lambda_{d}+2}}\Big )
\ee
if $\a_i + \lambda_i = -2$ for $i = 1,...,k$ and $a_i + \lambda_i \not= -2$
for $i = k+1,...,d$, where $0 \leq k \leq d-1$. Hence, we see that

\be
  {{{\rm d}V(t)}\over{{\rm d} t}} = \lim_{\v \to 0^{+}}
 {{{\rm d}V_{\v}(t)}\over{{\rm d} t}} \not = 0
\ee
which proves that shattering does indeed take place for $\a < 0$.

%=======================================================================

\section{Volume distributions}
\vspace{5mm}

In this section we briefly consider the volume distribution function
$C(V,t)$ defined by

\be
  C(V,t) = \int_{0}^{\infty}{\rm d}x_1...
    \int_{0}^{\infty}{\rm d}x_d \d (x_1...x_d - V) c(x_1,...,x_d,t) 
\ee
which is usually very useful for providing a partial description of a 
fragmenting system.

\vspace{5mm}

\begin{flushleft}
{\large\bf Model 1}
\vskip3mm
\end{flushleft}

For $\a_i = 0$, $ i = 1,...,d $, and mono-disperse initial conditions the 
relevant exact solution to this model is

\begin{eqnarray}
   c(x_1,...,x_d,t) &=& e^{-t}\Big [
      \d (x_1-l_1)...\d (x_d-l_d) + \nonumber  \\
     & &  {1\over{l_1...l_d}}\sum\limits_{r=0}^{\infty}
         {{(2^dt)}^{(r+1)}\over{(r+1)!(r!)^d}}
          \Bigg (\ln \Big ({l_1\over{x_1}}\Big )...
            \ln \Big ({l_d\over{x_d}}\Big )\Bigg)^r \Big ].
\end{eqnarray}
Substituting this into (71) yields

\be
   C(V,t) = e^{-t}\Big [
      \d (V-l_1...l_d) +
        {2^dt\over{l_1...l_d}}\sum\limits_{r=0}^{\infty}
         {{(2^dt)}^{r}\over{(r+1)![d(r+1)-1]!}}
       \Bigg (\ln \Big ({l_1...l_d\over{V}}\Big )\Bigg)^{[d(r+1)-1]} \Big ]. 
\ee
This reduces to the result of Ziff and McGrady [16] when $d = 1$, and to 
the result of Rodgers and Hassan [27] for $d = 2$. Expanding (73) for 
small $t$ gives

\be
   C(V,t) \sim  e^{-t}\Big [ \d (V-l_1...l_d) +
        {2^dt\over{l_1...l_d}} {1\over{(d-1)!}}
       \Bigg (\ln \Big ({l_1...l_d\over{V}}\Big )\Bigg)^{(d-1)} \Big ].
\ee
As $d$ increases by one, the power of the logarithmic divergence in the 
second term in (74) also increases by a factor of one. Analogous to the 
1-dimensional case, the d-dimensional case forms the borderline case for 
the shattering transition.

Define the normalized $n$th moments of the volume, $V(t)$, by 

\be
  <V^n> = {{\int_{0}^{\infty}}{\rm d}V V^n C(V,t)
             \over{\int_{0}^{\infty} {\rm d}V C(V,t)}}.
\ee
Then, it follows that

\be
   <V^n>^{1/n} \sim e^{-2^d (1-{1/{(n+1)^d}}){t/n}}
\ee
which indicates that $C(V,t)$ does not exhibit a scaling form.

\vspace{5mm}

\begin{flushleft}
{\large\bf Model 2}
\vskip3mm
\end{flushleft}

In this model, for mono-disperse initial conditions, the exact solution 
is given by (38). Substituting (38) into (71) yields

\be
  C(V,t) = \left\{ \begin{array}{r@{\quad\quad}l}
       e^{-{{{l_1^{\a_1}}...{l_d^{\a_d}}t}/{2^d}}}
          \Bigg (\d (V - l_1...l_d) +  \hfill &  \\
            \sum\limits_{r=1}^{\infty}  2^{{{r(r+1)\a}/{2}}+rd} 
             \d (V - {{l_1...l_d}\over{2^{rd}}})
           \sum\limits_{k=0}^{r}
          {{e^{{{l_1^{\a_1}}...{l_d^{\a_d}}(1-2^{(k-r)\a})t}/{2^d}}}
         \over{\P (2^{m \a}-2^{k \a})}} \Bigg ) &  {\rm ;} \a \not= 0   \\
     e^{-{{{l_1^{\a_1}}...{l_d^{\a_d}}t}/{2^d}}}
        \sum\limits_{r=0}^{\infty}{t^r\over{r!}}
         ({l_1^{\a_1}}...{l_d^{\a_d}})^r
             \d (V - {{l_1...l_d}\over{2^{rd}}})
                \hfill & {\rm ;}  \a = 0 \end{array} \right..
\ee
These results may be compared with the results when $d = 1$ for 
$c(x_1,...,x_d,t)$ given in (38) or in [30]. The similarity is quite 
remarkable, indicating that the scaling and shattering behaviour of 
$C(V,t)$ given in (77) will match closely that observed for 
$c(x_1,...,x_d,t)$ given by (38), when $d = 1$. 
%========================================================================

\section{Conclusions}
\vspace{5mm}

In reality, fragmenting particles will have both size and shape, i.e., a 
geometry. Intrigued by the possibility that the geometry of the 
fragmenting particles may influence the fragmentation process, we have 
investigated two distinct $d$-dimensional fragmentation models, for $d 
\geq 1$. Two classes of exact solutions to these geometric models, which 
describe the kinetics of fragmentation of $d$-dimensional particles, are 
presented. The first class is described by a fragmentation rate 
$a(x_1,...,x_d) = 1$ and a daughter distribution function $b(x_1,...,x_d 
| x_1^{\p},...,x_d^{\p}) = {{({\a_1} + 2)...({\a_d} + 
2){{x_1}^{\a_1}}...{{x_d}^{\a_d}}}
/{{{x_1}^{\p(\a_1 + 1)}}...{{x_d}^{\p(\a_d + 1)}}}}$. For $d > 1$, this 
particular class of exact solutions exhibits multi-scaling and does not 
permit the occurrence of a shattering transition. The second class of 
exact solutions is described by a
fragmentation rate $ a({x_1},...,{x_d}) = 
{{{x_1}^{\a_1}}...{{x_d}^{\a_d}}/{2^d}}$
and a daughter distribution function $b({x_1},..,{x_d} |
{{x_{1}^{\p}}},...,{{x_{d}^{\p}}}) = {2^d}{\d(x_1 -
{{x_{1}^\p}}/2)...\d(x_d - {{x_{d}^\p}}/2)}$. This particular class of exact
solutions describes a type of fragmentation process in which particles 
always break-up into $2^d$ equally sized pieces at various rates which 
depend upon the geometry of the fragmenting particles and the homogeneity 
indices $\a_1,...,\a_d$. this type of fragmentation has been observed and 
studied when polymers degrade under tension (stretching) [33], or in the 
presence of a destructive force-field such as ultra-sound [6]. Defining, 
$\a$, to be the sum of all the homogeneity indices $\a_1,...,\a_d$, it is 
shown that this particular class of exact solutions exhibits scaling for 
$\a > 0$ and this scaling form is explicitly determined. For $\a = 0$, we 
show that multi-scaling is likely and that the shattering transition is 
not permitted. When $\a < 0$, we show that a shattering transition occurs.

An interesting scenario occurs in our investigation into the scaling 
behaviour of the second class of exact solutions. When $\a > 0$, we have 
shown that a scaling solution to our model exists and we explicitly find 
this scaling solution. This is very surprising, since it has been 
suggested [28] that scaling solutions are not supposed to exist when we 
have an infinite number of hidden conserved integrals. We suggest that 
the existence of an infinite number of hidden conserved integrals is not 
always indicative of the absence of scaling solutions. There must exist 
other important criteria which conclusively indicate the absence of 
scaling in a particular $d$-dimensional fragmention process. We propose 
to investigate the nature of these criteria in subsequent work.

An investigation into the occurrence of a shattering transition in the second
class of exact solutions presented in section 2 is quite intriguing. when
$\a < 0$ it is shown that shattering occurs. It is interesting that whether
shattering occurs or not is determined exclusively by the fact that $\a < 0$,
and not on the sign of the individual $\a_1,...,\a_d$, which add up to 
give $\a$. Of course, the ferocity of the shattering transition will 
depend on how negative $\a$ is, with the possibility of competing effects 
between positive and negative $\a_i$, $ i = 1,...,d$, which will act to 
moderate the ferocity of this shattering transition.

The volume distribution function, $C(V,t)$, is very useful for providing 
a partial description of a fragmenting system. For small $t$ the 
2-dimensional model differs from the 1-dimensional model by the presence 
of a logarithmic divergence term. The 3-dimensional model differs from 
the 1-dimensional model by the presence of a logarithm squared divergence 
term, and so on. An analysis of the normalized $n$th momentsof the volume 
$V$ indicate that $C(V,t)$ does not exhibit scaling. For the second class 
of exact solutions presented in section 2, it is shown that the 
similarity between $C(V,t)$ and $c(x_1,...,x_d,t)$, with $d = 1$, is 
quite remarkable. This is a strong indicator that the scaling and 
shattering behaviour of $C(V,t)$ will be very similar to that observed 
for $c(x_1,...,x_d,t)$, with $d = 1$.

We have found that the introduction of more than one parameter to 
characterize the geometry of the fragmenting particle can have a 
significant effect on the kinetics of the fragmentation process. As a 
propect for future research, one could investigate problems with sources 
and sinks in $d$-dimensions, given that one now has a useful method 
available for determining exact solutions to geometric models which 
describes the kinetics of fragmentation of $d$-dimensional particles.

\pagebreak

%=======================================================================

\begin{flushleft}
{\large\bf Acknowledgement}
\vskip5mm

P. Singh would like to thank the EPSRC for financial support 
under grant GR/J25918 and M K Hassan would like the CVCP for ORS award.

\end{flushleft}
\pagebreak

%=======================================================================

\begin{flushleft}
{\large\bf References}
\vskip5mm

[1] R. Shinnar, J. Fluid Mech. {\bf 10}, 259 (1961).

[2] R. Meyer, K. E. Almin, and B. Steenberg,  Br. J. Appl. Phys. {\bf 17}, 
409 (1966). 

[3] W. R. Johnson and C. C. Price, J. Polym. Sci. {\bf 45}, 217 (1960).

[4] E. W. Merrill, H. S. Mickley and A. J. Ram, J. Polym. Sci. {\bf 62}, 
S109 (1962).

[5] A. M. Basedow, K. H. Ebert and H. J. Ederer, Macromolecules {\bf 
11}, 774 (1978).

[6] P. A. Glynn, B. M. van der Hoff and P. M. Reilly, J. Macromol. Sci. 
Chem. {\bf A6}, 1653, (1972).

[7] S. Egusa, K. Ishigure and Y. Tabata, Macromolecules {\bf 12}, 939 (1979).

[8] W. Kuhn, Ber. Chem. Dtch. Ges. {\bf 63}, 1503 (1930).

[9] H. Mark and R. Simha, Trans. Faraday Soc. {\bf 35}, 611 (1940).

[10] E. W. Montroll and R. Simha, J. Chem. Phys. {\bf 8}, 721 (1940).

[11] P. J. Blatz and A. V. Tobolsky, J. Chem. Phys. {\bf 49}, 77 (1945).
                                                                     
[12] H. H. Jellinek and G. White, J. Polym. Sci. {\bf 6}, 745 (1951).

[13] O. Saito, J. Phys. Soc. Jpn. {\bf 13}, 198 (1958).

[14] M. Ballauf and B. A. Wolf, Macromolecules {\bf 14}, 654 (1981).

[15] R. M. Ziff and E. D. McGrady, Macromolecules {\bf 19}, 2513 (1986).

[16] R. M. Ziff and E. D. McGrady, J. Phys. A: Math. Gen. {\bf 18}, 3027 
(1985).

[17] E. D. McGrady and R. M. Ziff, Phys. Rev. Lett. {\bf 58}, 892 (1987).

[18] R. M. Ziff, J. Phys. A: Math. Gen. {\bf 24}, 2821 (1991).

[19] T. A. Bak and K. Bak, Acta. Chem. Scand. {\bf 13}, 1997 (1959).

[20] A. Amemiya, J. Phys. Soc. Jpn. {\bf 17}, 1245 and 1694 (1962).

[21] M. Demjanenko and K. Du$\check{s}$ek, Macromolecules {\bf 13}, 571 
(1980).

[22] Z. Cheng and S. Redner, Phys. Rev. Soc. Lett. {\bf 60}, 2450 (1988).

[23] T. W. Peterson, Aerosol. Sci. Tech. {\bf 5}, 93 (1986).

[24] Z. Cheng and S. Redner, J. Phys. A: Math. Gen. {\bf 23}, 1233 (1990).

[25] M. H. Ernst G. Szamel, J. Phys. A: Math. Gen. {\bf 26}, 6085 (1993).

[26] A. F. Filippov, Theor. Prob. Appl. (USSR) {\bf 6}, 275 (1961).

[27] G. J. Rodgers and M. K. Hassan, Phys. Rev. E {\bf 50}, 3458 (1994).

[28] P. L. Krapivsky and E. Ben-Naim, Phys. Rev. E {\bf 50}, 3502 (1994).

[29] D. Boyer, G. Tarjus and P. Viot, Phys. Rev. E {\bf 51}, 1043 (1995).

[30] P. Singh and G. J. Rodgers, Phys. Rev. E {\bf 51}, 3731 (1995).

[31] R. M. Ziff, M. H. Ernst and E. M. Hendriks, J. Phys. A: Math. Gen. 
{\bf 16}, 2293 (1983).

[32] M. H. Ernst, R. M. Ziff and E. M. Hendriks, J. Coll. Int. Sci. {\bf 97},
266 (1984).

[33] A. Keller, J. A. Odell and H. H. Willis, Colloid. Polym. Sci. {\bf 263},
181 (1952).
\end{flushleft}
\end{document}